\documentclass[fleqn,usenatbib]{mnras}

\usepackage{newtxtext,newtxmath}

\usepackage[T1]{fontenc}

\DeclareRobustCommand{\VAN}[3]{#2}
\let\VANthebibliography\thebibliography
\def\thebibliography{\DeclareRobustCommand{\VAN}[3]{##3}\VANthebibliography}

\usepackage{graphicx}	
\usepackage{amsmath}	
\usepackage{siunitx}
\DeclareSIUnit{\bar}{bar}
\DeclareSIUnit{\year}{yr}
\DeclareSIUnit{\fl}{flashes}
\DeclareSIUnit{\nitrogen}{N}
\DeclareSIUnit{\lfr}{\fl\per\km\squared\per\year}
\DeclareSIUnit{\nflux}{\tera\g\nitrogen\per\year}
\usepackage[version=4]{mhchem}

\title[Lightning in storm-resolving simulations]{Lightning activity on a tidally locked terrestrial exoplanet in storm-resolving simulations for a range of surface pressures}

\author[D. E. Sergeev et al.]{%
Denis E. Sergeev,$^{1,2}$\thanks{E-mail: denis.sergeev@bristol.ac.uk}
James W. McDermott,$^{2,3}$
Lottie Woods,$^{2,4}$
Marrick Braam,$^{5}$
Jake K. Eager-Nash,$^{6,2}$
\newauthor
Ian A. Boutle$^{2,4}$
\\
$^{1}$School of Physics, University of Bristol, Bristol, BS8 1TL, UK\\
$^{2}$Department of Physics and Astronomy, University of Exeter, Exeter, EX4 4QF, UK\\
$^{3}$Centre for Life's Origin and Evolution, Department of Genetics, Evolution and Environment, University College London, London, WC1E 6BT, UK\\
$^{4}$Met Office, FitzRoy Road, Exeter, EX1 3PB, UK\\
$^{5}$Center for Space and Habitability, University of Bern, Gesellschaftsstrasse 6, 3012 Bern, Switzerland\\
$^{6}$School of Earth and Ocean Sciences, University of Victoria, Victoria, BC, V8P 5C2, Canada\\
}

\date{Accepted XXX. Received YYY; in original form ZZZ}

\pubyear{\the\year{}}

\begin{document}
\label{firstpage}
\pagerange{\pageref{firstpage}--\pageref{lastpage}}
\maketitle

\begin{abstract}
Cloudy atmospheres produce electric discharges, including lightning.
Lightning, in turn, provides sufficient energy to break down air molecules into reactive species and thereby affects the atmospheric composition.
The climate of tidally locked rocky exoplanets orbiting M-dwarf stars may have intense and highly localised thunderstorm activity associated with moist convection on their day side.
The distribution and structure of lightning-producing convective clouds is shaped by various climate parameters, of which a key one is atmospheric mass, i.e. surface air pressure.
In this study, we use a global storm-resolving climate model to predict thunderstorm occurrence for a tidally locked exoplanet over a range of surface pressures.
We compare two lightning parameterisations: one based on ice cloud microphysics and one based on the vertical extent of convective clouds.
We find that both parameterisations predict that the amount of lightning monotonically decreases with surface pressure due to a weaker convection and fewer ice clouds.
The spatial distribution of lightning on the planet changes with respect to the surface pressure, responding to the changes in the large-scale circulation and the vertical stratification of the atmosphere.
Our study provides revised, high-resolution estimates for lightning activity on a tidally locked Earth-like exoplanet, with implications for global atmospheric chemistry.
\end{abstract}

\begin{keywords}
Planets and satellites: terrestrial planets -- planets and satellites: atmospheres -- methods: numerical
\end{keywords}


\section{Introduction}
With the advent of powerful telescopes such as \textit{the James Webb Space Telescope (JWST)} and future facilities such as \textit{the Extremely Large Telescope (ELT)}, \textit{the Habitable Worlds Observatory (HWO)}, and \textit{the Large Interferometer For Exoplanets (LIFE)}, obtaining an atmospheric spectrum for an Earth-sized exoplanet and examining it for the presence of biomarker molecules will soon be within our reach.
For a robust interpretation of observational data, it is imperative to improve the theory of planetary atmospheres and consider all possible physical and chemical processes, including abiotic chemical pathways associated with atmospheric electric discharges, such as lightning.
Here, we focus on the lightning activity on a terrestrial exoplanet and examine the effect of varying atmospheric mass on the climatic precursors of lightning.

Lightning is a unique marker of convection and cloud microphysics on Earth and in extraterrestrial planetary atmospheres \citep[e.g.,][]{Helling19_lightning}.
Lightning has been detected in the cloudy atmospheres of solar system's giant planets, and tentatively on rocky bodies: Venus, Mars and Saturn's moon Titan \citep{Yair12_new, Hodosan16_lightning, Aplin17_lightning, Becker20_small}.
Observational evidence is mounting for the presence of clouds on exoplanets \citep[e.g.,][]{Kreidberg14_clouds, Sing16_continuum, Grant23_jwsttst, Kempton23_reflectivea}, so we can expect charging processes and electric storms to occur there too \citep{Helling10_ionization, Helling11_ionization, Helling13_ionization, Helling19_lightning, Hodosan21_exploring}.

Lightning is an important driver of disequilibrium chemistry, providing high-temperature channels up to \qty{30000}{\K} that allow for thermal decomposition of ambient atmospheric molecules into more reactive species.
As the heated gas cools, lightning-generated species are frozen out in disequilibrium abundances \citep[e.g.,][]{Chameides81_rates}.
The species produced by lightning vary with the atmospheric composition \citep{Chameides81_rates, NnaMvondo01_production, Barth24_effect}.
On Earth, lightning dissociates ambient \ce{N2} and \ce{O2} to form nitrogen oxides that affect abundances of oxidants and, as a result, air quality \citep{Schumann07_global, Murray16_lightning}.
Under more reducing Early Earth conditions, lightning produces prebiotic compounds \citep{Miller53_production, Cleaves08_reassessment}.
On terrestrial exoplanets, lightning is expected to similarly affect the atmospheric chemistry, both by providing pathways to prebiotic compounds \citep{Ardaseva17_lightning, Rimmer19_hydrogen, Pearce22_rna, Barth24_effect} and by affecting the abundances of potential biosignatures like ozone \citep{Ardaseva17_lightning, Harman18_abiotic, Braam22_lightning-induced}.
Crucially, future observations may be misinterpreted by overlooking important abiotic chemical pathways, such as lightning --- instead attributing markers of disequilibrium chemistry to biotic sources and falsely claiming the detection of life.


On Earth, lightning is predominantly generated in deep cumulonimbus clouds, which develop during moist convection \citep[e.g.,][]{Vonnegut63_facts, Williams85_largescale}.
These clouds attain large vertical depths and comprise mixed phase water particles, whose interaction leads to electric charging, followed by charge separation and eventual field breakdown in the form of lightning.
The most efficient non-inductive charging mechanism relies on the collisions of graupel particles with ice crystals in the presence of supercooled water droplets \citep{Saunders08_charge}.
The difference in fall speeds of these charged particles causes turbulent motions to separate them vertically and build a strong electric field within a thunderstorm.

Cloud distribution on terrestrial exoplanets presents a unique and fascinating scenario for lightning activity.
Earth-sized exoplanets orbiting M-dwarf stars are most favourable for atmospheric detection and characterization \citep[e.g.,][]{Dressing15_occurrence}.
Because M-dwarfs are cooler and dimmer compared to the Sun, planets around them that receive an Earth-like amount of stellar irradiation, are in close orbits.
Planets in close orbits are expected to be synchronous rotators due to tidal locking --- they have a permanent day and night hemispheres \citep{Joshi97_synchronous, Pierrehumbert19_atmospheric}.
Climate modelling suggests that the day side of these planets is enveloped in a thick layer of convective clouds \citep[e.g.,][]{Yang13_stabilizing, Leconte13_3d, Turbet16_habitability, Boutle17_exploring, Way18_climates, Komacek19_atmospheric, Paradise22_exoplasim, Wolf22_exocam}.
This cloud pattern may create a unique global electric field and favourable conditions for intense and locally concentrated lightning activity \citep{Helling19_lightning}.
Its strength depends on the intricacies of the cloud structure, which is in turn closely linked to the large-scale three-dimensional atmospheric dynamics \citep{Braam22_lightning-induced}.

Surface pressure, i.e. atmospheric mass per unit area, is one of the key unknowns for rocky exoplanets.
The bulk mass of an atmosphere is shaped by many processes such as planetary formation, atmospheric escape, impacts, volcanism and ocean chemistry \citep{Wordsworth22_atmospheres}.
Within our solar system, atmospheric mass on rocky bodies varies by orders of magnitude: e.g., from \qty{\sim 0.006}{bar} on Mars, to \qty{\sim 1}{bar} on Earth, to \qty{\sim 1.5}{bar} on Titan, to \qty{\sim 92}{bar} on Venus \citep[e.g.,][]{Showman13_atmospheric}.
Rocky exoplanets within the habitable zone around M-dwarf stars may have a wide range of atmospheric mass.
They may have either mostly lost their atmospheres to space via thermal escape \citep[e.g.,][]{Looveren24_airy, Looveren25_habitable} or retained them with a surface pressure of several bars and a large water inventory, as recent models of interior-atmosphere interaction suggest \citep{Krissansen-Totton24_erosion}.

The amount of atmospheric mass shapes the planetary climate through a number of radiative and thermodynamic processes, explored in the context of heat transport and climate transitions to a greenhouse state on asynchronously rotating terrestrial planets in e.g., \citet{Charnay13_exploring, Wolf14_delayed, Kaspi15_atmospheric, Chemke16_thermodynamic, Chemke17_dynamics}.
For tidally locked exoplanets, background air pressure affects the limits of their habitability \citep[e.g.,][]{Turbet18_modeling, Komacek19_atmospheric, Zhang20_background, Macdonald25_climate}.
It has been suggested that the instellation threshold for the runaway greenhouse state is a non-monotonous function of background pressure due to a competition between different climate feedbacks associated with temperature lapse rate, pressure broadening, water vapour and clouds \citep{Zhang20_background}.
Here, we explore how atmospheric mass affects the structure and distribution of convective lightning-generating clouds.
Our aim can be summarised in the following question: \textit{at which surface pressure does a tidally locked exoplanet have the most favourable climate for lightning production?}

While lightning climatology on Earth has been thoroughly investigated over the last few decades \citep[e.g.][]{Christian03_global, Murray16_lightning, Finney18_projected, Field18_simulated}, such studies for exoplanets are scarce and so far have been mostly based either on extrapolation of the Solar System data \citep{Hodosan16_lightning} or one-dimensional (1D) cloud-free modelling \citep{Ardaseva17_lightning, Harman18_abiotic}.
The full complexity of cloud formation and transport is unfeasible to simulate accurately in a 1D model.
The 3D aspect is crucial to correctly model the global thunderstorm climatology on an exoplanet as shown by \citet{Braam22_lightning-induced}; it is also key to capture the global transport of chemical species \citep[e.g.,][]{Yates20_ozone, Braam23_stratospheric, Cooke23_degenerate, Zamyatina24_quenchingdriven}.
At the same time, a large uncertainty in the thunderstorm climatology for the present \citep{Tost07_lightning, Stolz21_evaluating} and future \citep{Clark17_parameteriza, Finney18_projected} climate arises because moist convection is not fully resolved by coarse-grid 3D climate models.
Employing a high-resolution non-hydrostatic model, which simulates convection explicitly without relying on parameterizations, circumvents this problem \citep[e.g.,][]{Field18_simulated} and for this reason is used in this study.

The structure of this paper is as follows.
In Sec.~\ref{sec:methods}, we describe our general circulation model (Sec.\ref{sec:model}), its two lightning parameterisations (Sec.\ref{sec:lightning_param}), and our experimental setup (Sec.\ref{sec:experiments}).
In Sec.~\ref{sec:results}, we analyse the simulated lightning climatology and its driving mechanisms, showing that the amount of lightning decreases with increasing surface pressure, as predicted by both the microphysics-based scheme (Sec.~\ref{sec:res_m09}) and cloud depth-based scheme (Sec.~\ref{sec:res_pr92}).
In Sec.~\ref{sec:discussion}, we discuss the uncertainties and caveats in our study, as well as its implications for atmospheric chemistry on tidally locked planets.
In Sec.~\ref{sec:conclusions}, we summarise our findings.

\section{Methods}
\label{sec:methods}
We simulate cloud-generated lightning activity, assuming a tidally locked aquaplanet with a nitrogen-dominated atmosphere with a range of total surface pressures.
We use a 3D general circulation model, the Met Office Unified Model (UM).
While used primarily for Earth climate and weather prediction \citep{Walters19_ga7, Andrews20_historical}, the UM has been applied to study atmospheric processes on a variety of rocky planets, including an idealised Earth-like exoplanet \citep{Mayne14_idealised}, Archean Earth \citep{Eager-Nash23_3d, Mak23_3d}, Mars \citep{McCulloch23_modern}, Proxima Centauri b \citep{Boutle17_exploring, Lewis18_influence, Joshi20_earth, Sergeev20_atmospheric, Boutle20_mineral, Yates20_ozone, Braam22_lightning-induced, Cohen22_laso, Ridgway22_3d, Braam23_stratospheric, Cohen23_traveling, Braam25_earthlike} and TRAPPIST-1e \citep{Eager-Nash20_implications, Sergeev20_atmospheric, Turbet22_thai, Sergeev22_thai, Sergeev22_bistability, Cohen23_traveling, Mak24_3d}.
In this project, we use the UM in a global storm-resolving configuration \citep{Field18_simulated} applied to an extraterrestrial atmosphere for the first time.

\subsection{Model description}
\label{sec:model}
All simulations in this study are performed with the UM \texttt{vn12.2} in the configuration close to the GA7.0 science configuration \citep{Walters19_ga7}.
It includes parameterizations of large-scale cloud \citep{Wilson08_pc2} and cloud microphysics \citep{Wilson99_microphysically}, convection \citep{Gregory90_mass} and lightning (see Sec.~\ref{sec:lightning_param}).

The radiative transfer in the UM is computed by the Suite of Community Radiative Transfer codes based on Edwards and Slingo (SOCRATES) scheme.
This scheme uses the correlated-k method and has been applied for a range of non-Earth configurations \citep{Amundsen14_accuracy, Amundsen16_um} and is currently being benchmarked in the radiative transfer code intercomparison for exoplanets \citep{Villanueva24_modeling}.
Opacity data in the form of `spectral files' for SOCRATES were obtained from the NASA Goddard Institute for Space Studies\footnote{\url{https://portal.nccs.nasa.gov/GISS_modelE/ROCKE-3D/spectral_files}}.

Following \citet{Field18_simulated}, we use the UM in the explicit convection configuration (i.e., the convection scheme is turned off) with the N1280 grid.
This high-resolution grid has a spacing of \ang{0.140625} in longitude and \ang{0.09375} in latitude, corresponding to a median grid spacing of \qty{\approx 10}{\km} for TRAPPIST-1e (Table~\ref{tab:exp}).
In the vertical, we use 63 vertical levels between the surface and the model top, located at a height of \qty{\approx 40}{\km}.
The N1280 model is initialised from a steady-state climate, achieved by running the UM at the N96 resolution.
N96 is one of the standard grid resolutions (\ang{1.875}$\times$\ang{1.25}) used by the Met Office for climate simulations.
The steady state is assumed when the top-of-atmosphere (TOA) radiation balance stabilised close to zero and the global mean surface temperature reached a quasi-constant value.
For consistency, we use 3600~days ($\approx$590 TRAPPIST-1e orbits) as the spin-up period for all simulations, even though low-pressure simulations reach the steady state earlier.
We then run the model in the high-resolution configuration for 365 days and analyse the results (Sec.~\ref{sec:res_m09}) averaged over 5 subsequent days (days 366--370).
For the low-resolution simulations used in Sec.~\ref{sec:res_pr92}, we average the data over 30 days.

\begin{table}
    \centering
    \caption{The UM configurations used in this study.
    Grid spacing in the sixth row is calculated as the square root of the
median grid cell area $\sqrt{A_{m}}$.
    \label{tab:exp}}
    {\scriptsize%
    \begin{tabular}{lll}
        \hline
         & M09 & PR92 \\
        \hline
        Lightning scheme & \citep{McCaul09_forecasting} & \citep{Price92_simple} \\
        Proxy for lightning & Ice cloud microphysics & Convective cloud depth \\
        Convection scheme & No & Yes \citep{Gregory90_mass} \\
        Prognostic graupel & Yes & No \\
        UM resolution & N1280 & N96 \\
        Longitudes $\times$ latitudes & 2560$\times$1920 & 192$\times$144 \\
        Grid spacing (\unit{\km}) & 9.8 & 130.4 \\
        Model time step (\unit{\s}) & 240 & 1200 \\
        \hline
    \end{tabular}
    }
\end{table}

\subsection{Lightning parameterisations}
\label{sec:lightning_param}
Simulating cloud electrification directly is computationally prohibitive in climate models; instead its occurrence is inferred from bulk cloud properties, such as the total amount or vertical flux of frozen particles, or the vertical extent of clouds \citep{Tost07_lightning, Stolz21_evaluating}.
In this study, we use two different lightning parameterisations: based on i) ice cloud microphysics (Sec.~\ref{sec:param_m09}) and ii) convective cloud depth (Sec.~\ref{sec:param_pr92}).

\subsubsection{Microphysics-based parameterisation (M09)}
\label{sec:param_m09}
In our storm-resolving simulations, lightning is diagnosed by the \citet{McCaul09_forecasting} scheme (hereafter M09) coupled to the Unified Model microphysics \citep{Wilkinson17_technique}, which include prognostic graupel (precipitating ice particles).
The lightning occurrence is parameterised to be \qty{95}{\percent} due to the upward flux of graupel and \qty{5}{\percent} due to a larger, widespread production of lightning typically in ice-dominated cloud anvils.
The total lightning flash rate is a combination of both factors:
\begin{equation}
    F = 0.95F_1 + 0.05F_2.
\end{equation}
Here, $F_1$ is a linear function of the upward flux of graupel mass at the \qty{-15}{\degreeCelsius} isotherm (where the cloud is assumed to be mixed-phase)
\begin{equation}
    F_1 = k_1 w q_g|_{-15C},
\end{equation}
where $w$ is the vertical velocity in \unit{\m\per\s} and $q_g$ is the mass mixing ratio of graupel in \unit{\g\per\kg}.
The $F_2$ factor is a linear function of the total ice water path (TIWP):
\begin{equation}
    F_2 = k_2 TIWP \equiv k_2 \int_0^{TOA}\rho(q_i + q_g) \mathrm{d}z,
\end{equation}
where $\rho$ is air density.
TIWP (\unit{\kg\per\m\squared}) is the vertically integrated mass of cloud ice and snow ($q_i$) and graupel ($q_g$).

The coefficients $k_1$ and $k_2$ are determined empirically from observations and available as inputs to the scheme.
\citet{McCaul09_forecasting} used the values $k_1=0.042$ and $k_2=0.21$, while here we used the updated values of 0.21 and 0.6, respectively (following operational guidance from the Met Office).
The exact values of $k_1$ and $k_2$ are naturally a source of uncertainty for modelling lightning activity for any atmosphere so in our study we focus on the overall trends rather than absolute numbers.

This parameterisation was originally derived from observations of severe thunderstorms over the United States and thus may suffer from a regional bias.
In the UM, it has been validated against regional weather forecasts and global climatology \citep{Wilkinson17_technique, Field18_simulated}.
While the spatial distribution and seasonal cycle are captured well, this scheme tends to overestimate the lightning flash rate over the oceans \citep{Field18_simulated}. 



\subsubsection{Cloud depth-based parameterisation (PR92)}
\label{sec:param_pr92}
In coarse-grid models with parameterised convection, lightning occurrence is often diagnosed by the vertical extent of convective clouds \citep{Tost07_lightning}.
Conceptually, these schemes relate cloud top height ($H$, in \unit{\km}) to the number of flashes.

Summarising the theory from \citet[][hereafter PR92]{Price92_simple}, we demonstrate that for a thunderstorm, i.e., a region of space charge $\rho_c$, the electrical energy $W$ increases with its height \citep[see also e.g.,][]{Vonnegut63_facts, Williams85_largescale}.
A thunderstorm has regions of positive and negative charge ($Q$ and $Q'$) which create the electrical power $P$:
\begin{equation}
    P = K\frac{QQ'}{R},
\end{equation}
where $R$ is the distance between the charges and $K=\frac{1}{4\pi\epsilon_0}$.
Given that $Q$ and $Q'$ are equal to the charge density $\rho_c$ multiplied by the volume that is proportional to $H^3$, and that $R\sim H$, we obtain
\begin{equation}
    P \sim K\rho_c^2H^5.
\end{equation}
Since the electrical power $P$ is the rate of generation of electrical energy ($P=W/t$, where $t$ is time), an increase in storm height (and therefore electrical power) implies an increased rate of electrical energy generation $W$ and hence more frequent lightning.

In the UM, the cloud-depth based lightning scheme follows the original work in \citet{Price92_simple}, with further implementation details given in \citet{Allen02_evaluation}.
The scheme takes the base and top convective cloud levels from the mass flux convection scheme \citep{Gregory90_mass} and calculates the convective cloud depth.
If the convective cloud depth exceeds \qty{5}{\km}, the cloud top height is used to calculate the flash rate as
\begin{equation}
    F = \mathcal{A} H^\mathcal{B},
\end{equation}
where $\mathcal{A}$ and $\mathcal{B}$ are empirical constants that are different for land and ocean grid points.
In this study, we assume an aquaplanet (see Sec.~\ref{sec:experiments}) so only the oceanic relationship is used, with the empirical constants from \citet{Luhar21_assessing}:
\begin{equation}
    F_{ocean} = 2\times10^{-5} H^{4.38}.
\end{equation}

While the \citet{Price92_simple} scheme was developed for Earth, we justify its use in our study by assuming an Earth-like atmosphere with moist convection and therefore, a similar physical process responsible for the charge separation.
In Sec.~\ref{sec:discussion}, we discuss the applicability of this scheme for exoplanets.
We apply the \citet{Price92_simple} scheme to the N96 (coarse-resolution, see Table~\ref{tab:exp}) UM simulations with parameterised convection; this is why this scheme is sometimes called convection-based. 
Note that the same scheme was used in coarse-grid (N72) chemistry-climate simulations for Proxima Centauri b by \citet{Braam22_lightning-induced}, providing us with an additional point of comparison for a similar planetary setup.


\subsection{Experiments}
\label{sec:experiments}


\begin{table}
	\centering
	\caption{Stellar spectrum and planetary parameters for TRAPPIST-1e following \citep{Fauchez20_thai_protocol}.\label{tab:planet}}
	\begin{tabular}{lll} 
		\hline
Parameter         & Value                              & Units \\
        \hline
Star and spectrum & \qty{2600}{\K} BT-Settl with Fe/H=0 & \\
Semi-major axis   & 0.02928                            & AU \\
Orbital period    & 6.1                                & Earth day \\
Rotation period   & 6.1                                & Earth day \\
Obliquity         & 0                                  & degrees \\
Eccentricity      & 0                                  & \\
Instellation      & 900.0                              & \unit{\watt\per\square\meter} \\
Planet radius     & 5798                               & \unit{\km} \\
Gravity           & 9.12                               &  \unit{\meter\per\second\squared}\\
		\hline
	\end{tabular}
\end{table}

Our simulations are configured with the planetary parameters from the TRAPPIST-1e Habitable Atmosphere Intercomparison (THAI) protocol \citep[][see also Table~\ref{tab:planet}]{Fauchez20_thai_protocol}.
The planet is assumed to be in 1:1 synchronous rotation, i.e. its rotation period is equal to its orbital period; obliquity and eccentricity are assumed to be zero.
The assumption of the 1:1 synchronous rotation is justified by the likely time scale of tidal locking of TRAPPIST-1e compared to the age of its host star \citep{Barnes17_tidal, Turbet18_modeling}.

We set our model with \ce{N2} as the dominant background gas and 400\,ppm of \ce{CO2} (i.e., fixed mass mixing ratio), and \ce{H2O} as the condensible species.
Oxygen and ozone are not included in our simulations.
While an \ce{O2}-dominated atmosphere is a candidate for the composition of TRAPPIST-1e \citep{Turbet20_review}, the bulk radiative forcing of \ce{O2} is broadly similar to that of \ce{N2} \citep{Fauchez19_impact} and its presence primarily affects the chemical \textit{impact} of lightning which is beyond the scope of this study \citep[see e.g.,][]{Chameides81_rates, Barth24_effect}.

For studying the effects of surface pressure on the lightning activity, we conduct 6 simulations with the surface pressures of 0.25, 0.5, 1, 2, 4, and 10\,bar, similar to e.g., \cite{Zhang20_background} and \citet{Macdonald25_climate}.
The 1\,bar simulation is equivalent to the Hab~1 case in the THAI project \citep{Fauchez20_thai_protocol, Sergeev22_thai}.
As mentioned above, each of these scenarios were modelled using i) a low-resolution UM with the PR92 lightning scheme and ii) high-resolution UM with the M09 lightning scheme.
The differences in the predicted climate between the two model resolutions are sufficiently small to facilitate this comparison.

\section{Results} \label{sec:results}
\begin{figure*}
    \centering
    \includegraphics[width=1\linewidth]{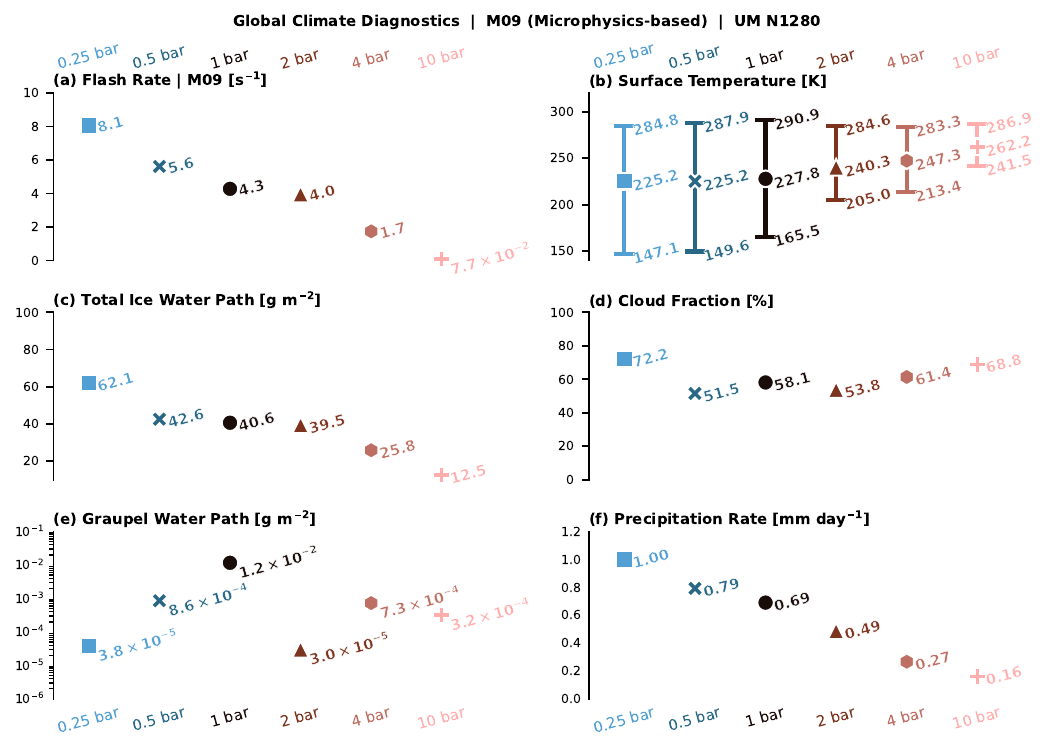}
    \caption{Global climate diagnostics in the high-resolution UM simulations with the surface pressure of 0.25, 0.5, 1, 2, 4, 10\,bar: (a) total flash rate in \unit{\per\s}, (b) surface temperature in \unit{\K}, (c) total ice water path in \unit{\kg\per\m\squared}, (d) cloud fraction in \unit{\percent}, (e) graupel water path in \unit{\kg\per\m\squared} and (f) precipitation rate in \unit{\mm\per\day}.
    Panel (b) also shows the minimum and maximum surface temperature.}
    \label{fig:glob_diag}
\end{figure*}

\begin{figure*}
    \centering
    \includegraphics[width=1\linewidth]{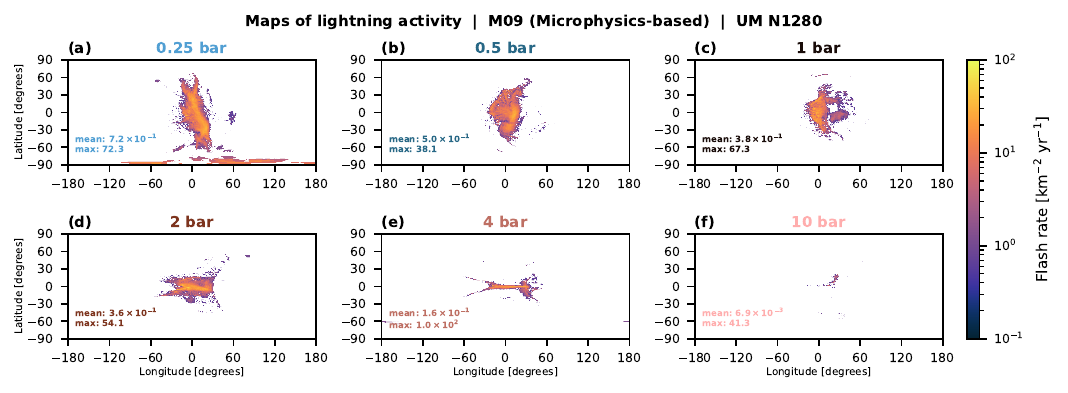}
    \caption{Maps of the lightning flash rate in \unit{\lfr} diagnosed by the M09 scheme in high-resolution UM simulations with the surface pressure of 0.25, 0.5, 1, 2, 4, 10\,bar.
    Note the logarithmic scale of the colour bar.}
    \label{fig:lfr_m09_maps}
\end{figure*}

\begin{figure*}
    \centering
    \includegraphics[width=1\linewidth]{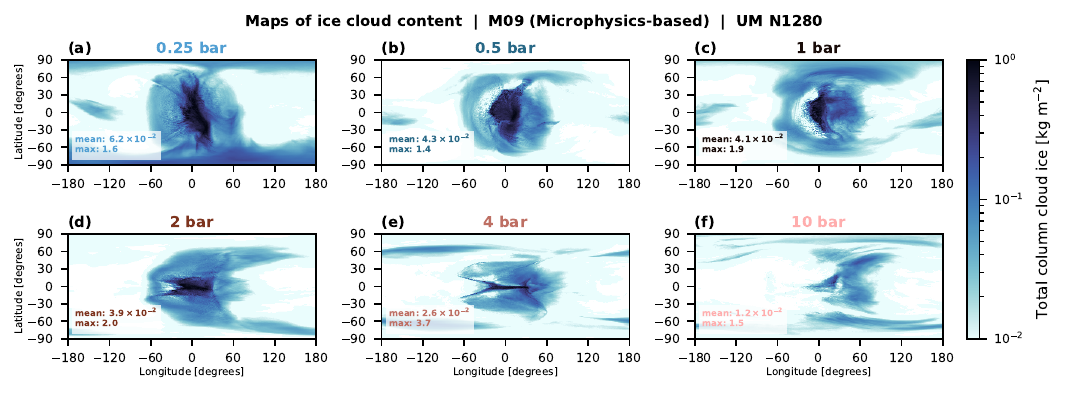}
    \caption{Maps of the total column cloud ice content in \unit{\kg\per\m\squared} in the high-resolution UM simulations with the surface pressure of 0.25, 0.5, 1, 2, 4, 10\,bar.}
    \label{fig:iwp_m09_maps}
\end{figure*}

\begin{figure*}
    \centering
    \includegraphics[width=1\linewidth]{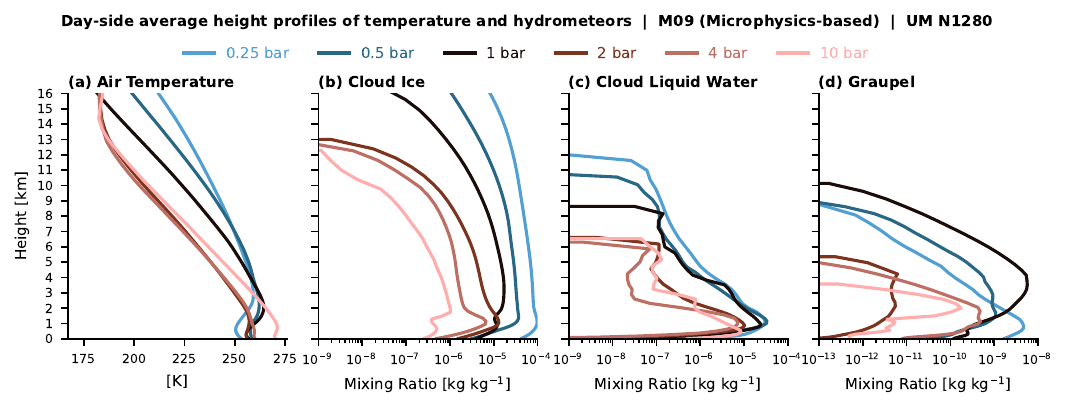}
    \caption{Day-side average (a) air temperature, (b) cloud ice, (c) cloud liquid water, and (d) graupel as a function of height in our M09 simulations for the surface pressure of 0.25, 0.5, 1, 2, 4, 10\,bar.
    The figure shows that with increasing surface pressure, the lapse rate becomes larger, the cloud water content diminishes, while the graupel content does not show a clear trend.
    Note that because the dry atmospheric mass differs between our simulations, the increase in the mass mixing ratios with pressure in this figure appears much greater than the increase in the vertically integrated cloud content shown in Figs.~\ref{fig:glob_diag}c and \ref{fig:iwp_m09_maps}.}
    \label{fig:vert}
\end{figure*}

\subsection{Monotonic decrease of lightning rate with surface pressure}
\label{sec:res_m09}
Our results show that the lightning flash rate (LFR) decreases monotonically with surface pressure.
This can be explained by the decrease of cloud thickness and cloud ice content, which are both consequences of weaker convection in high-pressure atmospheres.

The total global LFR in the control case (1~bar) simulation is \qty{4.3}{\per\s}, as diagnosed by the microphysics-based parameterisation M09 (Fig.~\ref{fig:glob_diag}a).
When the mass of the atmosphere is four times smaller (0.25~bar), LFR almost doubles to \qty{8.1}{\per\s}, whilst for a four times more massive atmosphere the rate is \qty{1.7}{\per\s}, dropping dramatically to \qty{7.7e-2}{\per\s} for a 10~bar atmosphere.
In the context of the global average LFR of \qty{44\pm5}{\per\s} observed on Earth \citep{Christian03_global}, our simulations produce much less lightning because the day-side surface is substantially cooler than the terrestrial tropical land regions, resulting in generally weaker convection \citep{Williams02_physical}.

As previous simulations for tidally locked planets suggest  \citep{Braam22_lightning-induced}, thunderstorms are concentrated on the planet's day side where the most intense convection happens, approximately within \ang{\pm 45} from the substellar point (Fig.~\ref{fig:lfr_m09_maps}).
The spatial distribution of lightning correlates with the maximum in the ice cloud water content (Fig.~\ref{fig:iwp_m09_maps}), which is the highest for the deep convection regions near the substellar point.
In the 1~bar case, the maximum LFR reaches \qty{67}{\lfr}, comparable to the highest LFRs over the ocean areas on Earth \citep{Christian03_global, Han21_cloud}.
From the lower to higher surface pressure, the area of high LFR gradually shrinks and shifts eastward.
This is due to the reduction in the ice clouds as well as the change in the global circulation.
Namely, the low-pressure simulations (\qtyrange{0.25}{1}{\bar}) favour the `single jet' \citep{Sergeev22_bistability}, also known as the `slow rotator' regime \citep{Haqq-Misra18_demarcating}, while the high-pressure simulations (\qtyrange{2}{10}{\bar}) favour the `double jet' or `Rhines rotator' regime.
In the former, the day-side cloud cover extends to higher latitudes, while in the latter, it is more zonally oriented and confined to low latitudes.
Consequently, the LFR field is concentrated near the equator in the high-pressure experiments, even though its average values are lower than those in the low-pressure experiments (Fig.~\ref{fig:lfr_m09_maps}).

The overall inverse proportionality of LFR with surface pressure in our model is explained by the amount of ice clouds.
To the first order, the amount of ice is proportional to the height of convective clouds, which become more and more dominated by the ice phase as the surface pressure decreases (Fig.~\ref{fig:vert}b,c).
As can be seen in the maps of vertically integrated cloud ice (Fig.~\ref{fig:iwp_m09_maps}), its global trend is downward with respect to the surface pressure, from \qty{62}{\g\per\m\squared} in the 0.25~bar simulation to \qty{40}{\g\per\m\squared} in the 1~bar simulation, to \qty{12}{\g\per\m\squared} in the 10~bar simulation (see also Fig.~\ref{fig:glob_diag}c).
Note that the amount of graupel, another key ingredient of the M09 lightning scheme, affects LFR in our simulations to a much lesser extent.
This is simply because of its low concentration, both in terms of the vertically integrated amount (Fig.~\ref{fig:glob_diag}e) and its vertical profile (Fig.~\ref{fig:vert}d).

\begin{figure}
    \centering
    \includegraphics[width=1\linewidth]{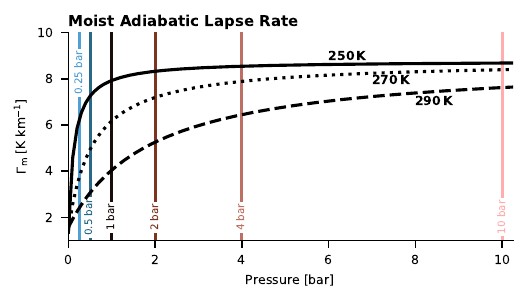}
    \caption{The increase of the moist adiabatic lapse rate $\Gamma_\mathrm{m}$ (vertical axis) with respect to pressure (horizontal axis) following Eq.~\ref{eq:gamma_m}.
    Additionally, $\Gamma_\mathrm{m}$ decreases with temperature as shown by the three different curves for a range of day-side values in our simulations: solid for \qty{250}{\K}, dotted for \qty{270}{\K} and dashed for \qty{290}{\K}.}
    \label{fig:gamma_m}
\end{figure}

The decrease of ice clouds with the surface pressure is driven by the three key climate mechanisms.
First, higher pressure suppresses convection by increasing the moist adiabatic (pseudoadiabatic) lapse rate \citep{Goldblatt09_nitrogenenhanced, Wolf15_evolution, Chemke16_thermodynamic, Zhang20_background}.
The lapse rate is the change in temperature with height, $\Gamma=-\mathrm{d}T/\mathrm{d}z$.
When the atmospheric lapse rate $\Gamma$ exceeds the moist adiabatic lapse rate $\Gamma_\mathrm{m}$, convection occurs and decreases $\Gamma$ until it equals $\Gamma_\mathrm{m}$ \citep[see e.g.,][]{Goldblatt09_nitrogenenhanced}.
The moist adiabatic lapse rate is defined as
\begin{equation}
    \Gamma_\mathrm{m} = \frac{g}{c_\mathrm{pd}}\left(\frac{1+\frac{L_\mathrm{v} q_\mathrm{sat}}{R_\mathrm{d} T}}{1+\frac{L_\mathrm{v}^2 q_\mathrm{sat} \epsilon}{c_\mathrm{pd}R_\mathrm{d} T^2}}\right),  \label{eq:gamma_m}
\end{equation}
where the saturation mixing ratio is
\begin{equation}
    q_\mathrm{sat} = \frac{\epsilon e_\mathrm{sat}}{p-e_\mathrm{sat}}, \label{eq:qsat}
\end{equation}
where the saturation vapour pressure is approximated by
\begin{equation}
    e_\mathrm{sat} = e_\mathrm{sat,tp} \mathrm{exp}\left[\frac{L_\mathrm{v}}{R_\mathrm{v}}\left(\frac{1}{T_\mathrm{tp}}-\frac{1}{T}\right)\right].  \label{eq:esat}
\end{equation}
In the equations above, $g$ is the acceleration due to gravity, $c_\mathrm{pd}$ is the specific heat capacity of dry air at constant pressure, $L_v$ is the latent heat of vaporisation for water, $R_\mathrm{d}$ is the dry air gas constant, $\epsilon$ us the ratio of molar masses of water and dry air, $T$ is temperature, $p$ is pressure, while $T_\mathrm{tp}$ and $e_\mathrm{sat,tp}$ are the temperature and saturation vapour pressure at the triple point of water, respectively.
Crucially, $\Gamma_\mathrm{m}$ depends on pressure (Eq.~\ref{eq:qsat}): increasing pressure increases the moist adiabatic lapse rate, therefore decreasing the amount of convection (Fig.~\ref{fig:gamma_m}).
Without an efficient convective energy flux, the lower troposphere heats up, while the upper troposphere cools down (Fig.~\ref{fig:vert}a), in agreement with predictions of 1D radiative-convective climate models \citep{Goldblatt09_nitrogenenhanced, Zhang20_background}.
Weaker convection results in shallower convective clouds, which do not develop ice-dominated anvils, which decreases the amount of ice particles needed for lightning generation.

Second, the greenhouse effect becomes stronger with higher atmospheric mass because of a stronger nitrogen gas pressure broadening \citep{Xiong20_possible, Zhang20_background} and a larger partial pressure of \ce{CO2} in our simulations.
Additionally, the Clausius-Clapeyron law creates a positive feedback loop of more water vapour being added to the atmosphere which in turn increases the greenhouse effect.
As a result, the lower troposphere becomes warmer (Fig.~\ref{fig:vert}a) and thus more liquid than frozen particles form in convective clouds.
The latter can be seen by the clear decrease in the cloud ice mixing ratio with pressure (Fig.~\ref{fig:vert}b), while cloud liquid water mixing ratio does not decrease at the same rate (Fig.~\ref{fig:vert}c).
Note that these differences hold when the total atmospheric mass is accounted for, i.e. when the cloud mass is integrated vertically in Fig.~\ref{fig:iwp_m09_maps}.

Third, for higher surface pressures, the night side warms up drastically (minimum surface temperature increases by almost \qty{100}{\K}, see Fig.~\ref{fig:glob_diag}b), decreasing the `radiator fin' effect and effectively raising the global mean temperatures \citep[for similar scenarios, see e.g.,][]{Turbet18_modeling, Macdonald25_climate}.
This is because the global zonal circulation switches to the double-jet regime \citep{Sergeev22_bistability}, and because the global overturning circulation --- upwelling on the day side and downwelling on the night side --- transports more energy from the day side to the night side in high-surface pressure simulations due to the overall higher mass of the atmosphere \citep{Kaspi15_atmospheric, Chemke17_dynamics}.

A counteracting effect that could be expected to reduce the LFR in the low-pressure climates is that they are colder and drier than the high-pressure climates.
However, the maximum surface temperatures in the low-pressure cases (\qtyrange{0.25}{1}{\bar}) are almost the same (Fig.~\ref{fig:glob_diag}b), suggesting that the moisture supply driven by the day-side surface evaporation is similar to that in the high-pressure cases and does not inhibit moist convection.
The total cloudiness does not show a clear trend either, hovering between \qtyrange{\approx 50}{70}{\percent} (Fig.~\ref{fig:glob_diag}d), and thus does not explain the LFR decrease with pressure.

Finally, while there is no clear trend of the graupel content with respect to the surface pressure, the total precipitation decreases monotonically (Fig.~\ref{fig:glob_diag}f), which agrees with the predicted downward trend of LFR.
The global mean precipitation rate drops from \qty{1}{\mm\per\day} in the 0.25~bar case to \qty{\approx 0.7}{\mm\per\day} in the control (1~bar) case to \qty{\approx 0.2}{\mm\per\day} in the 10~bar case.
This inverse relationship can be explained from the point of view of the energy balance of the atmosphere \citep[e.g.,][]{OGorman12_energetic}.
At equilibrium, precipitation (multiplied by the latent heat of condensation) is balanced principally by the difference between the net thermal (longwave) emission and the absorbed stellar (shortwave) radiation.
As the atmospheric mass increases, the net longwave emission decreases (stronger greenhouse effect), while the shortwave absorption increases, mainly by water vapour.
The smaller difference in these terms results in the decrease of precipitation in the climates with higher surface pressure, as discussed in detail in  \citet{Xiong22_smaller}.

\subsection{Lower lightning rate in the cloud-depth based scheme}
\label{sec:res_pr92}
\begin{figure*}
    \centering
    \includegraphics[width=1\linewidth]{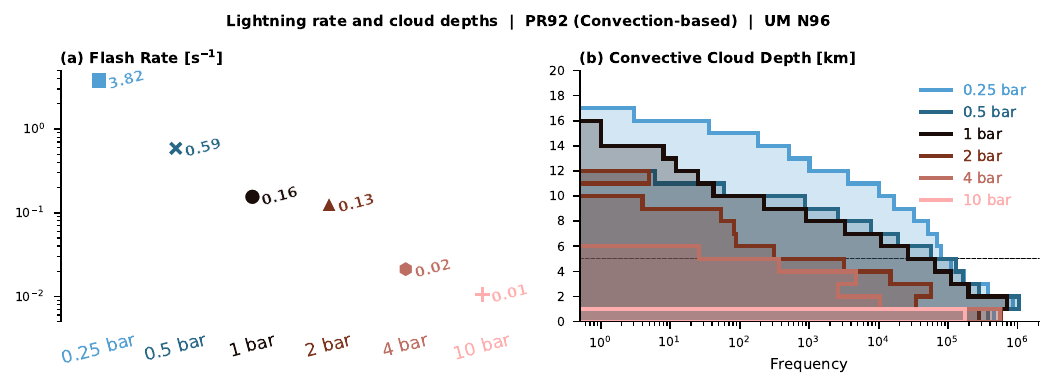}
    \caption{(a) Global flash rate (\unit{\per\s}) diagnosed by from the depth of convective clouds by the PR92 scheme.
    Note the logarithmic scale when comparing to Fig.~\ref{fig:glob_diag}a.
    (b) Frequency of convective clouds with different depth calculated using hourly data over 30 days of the n96 simulation.
    Cloud depths are calculated as the difference between the cloud top and cloud base.
    Different colour shadings correspond to different simulations.
    Convective clouds with depths \qty{> 5}{\km} (dashed black line) are classified as thunderclouds by the PR92 scheme (see Sec.~\ref{sec:param_pr92}).}
    \label{fig:glob_diag_pr92}
\end{figure*}

Compared to the microphysics-based (M09) estimates in Sec.~\ref{sec:res_m09}, the cloud-depth based (PR92) scheme consistently predicts less lightning.
As mentioned above, this is not due the climatic differences between the low- and high-resolution simulations --- these differences are very small (mean surface temperature differences are \qtyrange{\sim 1}{2}{\K}). 
As Fig.~\ref{fig:glob_diag_pr92}a shows, the global LFR in the reference case (1~bar) is estimated by the PR92 scheme to be \qty{0.16}{\per\s}, which is more than an order of magnitude smaller than the M09 estimate.
This difference broadly persists throughout our six scenarios, though it is the largest for the mid-range surface pressures.
For the 0.25~bar scenario, the inter-scheme difference is only of a factor of 2.
Part of this difference between the schemes may be explained by the known positive bias of the M09 scheme over the Earth oceans \citep{Field18_simulated}.

Despite the PR92 parametrization producing substantially less frequent lightning than M09, it exhibits the same monotonic downward trend of LFR with surface pressure.
This is linked to the vertical extent of deep convective clouds.
The prevalence of convective clouds of varying depth is shown in Fig.~\ref{fig:glob_diag_pr92}b.
This histogram of hourly data from our model clearly shows that convective clouds become shallower with increased atmospheric mass, due to the suppression of convection explained in Sec.~\ref{sec:res_m09}.
In the 0.25~bar case, the convective clouds grow to up to \qty{17}{\km} in thickness, while in the 10~bar case they are typically only \qty{1}{\km} thick.
In the 1~bar case, convective cloud depth reaches \qty{16}{\km} which agrees with the estimates for a similar climate simulated for Proxima Centauri b \citep[see Fig.~4 in][]{Braam22_lightning-induced}.
Note that in the 0.5~bar case, even though the cloud depth reaches only \qty{12}{\km} in our histogram, the LFR is larger than that in the 1~bar case.
This is because convective cloud tops reach high altitudes more frequently in the 0.5~bar case than in the 1~bar case (not shown).
While the convective cloud depth exceeding the \qty{5}{\km} threshold is the necessary condition for the cloud to be classified as lightning-producing in the PR92 scheme, it is the cloud top height that defines the flash rate  ($H^{4.38}$, see Sec.~\ref{sec:param_pr92}).

\begin{figure*}
    \centering
    \includegraphics[width=1\linewidth]{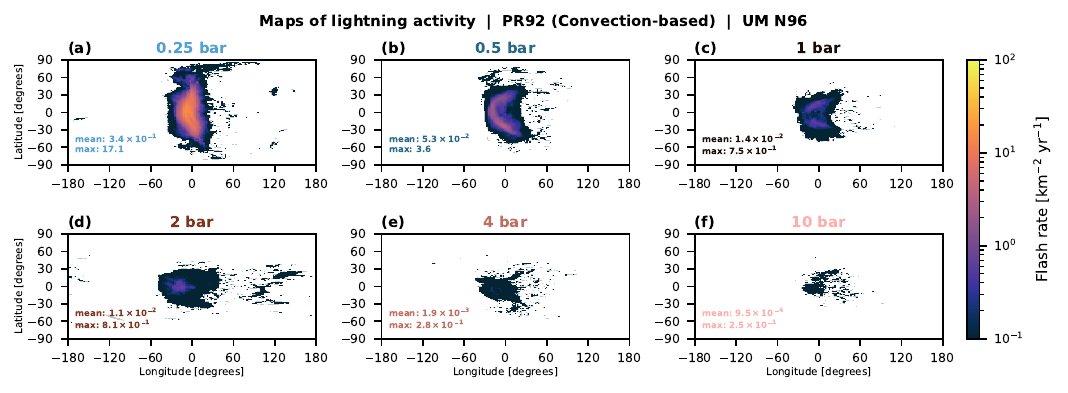}
    \caption{Maps of the lightning flash rate in \unit{\lfr} diagnosed by the PR92 scheme in low-resolution UM simulations with the surface pressure of 0.25, 0.5, 1, 2, 4, 10\,bar.}
    \label{fig:lfr_pr92_maps}
\end{figure*}

The geographical distribution of LFR estimated by the PR92 scheme (Fig.~\ref{fig:lfr_pr92_maps}) agrees well with that estimated by the M09 scheme.
In the 1~bar case, LFR is concentrated in a westward-facing crescent shape around the substellar point, corresponding to the deepest and persistent convection area.
Our results qualitatively agree with the Proxima Centauri b simulations \citep[see Fig.~3 in][]{Braam22_lightning-induced}, although LFR in our simulations is higher, reaching \qty{0.75}{\lfr}.
Maximum LFR decreases monotonically with atmospheric mass (with the exception of the 2~bar case), from \qty{17}{\lfr} in the 0.25~bar simulation to \qty{0.25}{\lfr} in the 10~bar simulation.

Both Fig.~\ref{fig:lfr_m09_maps} and Fig.~\ref{fig:lfr_pr92_maps} also show that weak lightning activity may occur on the night side of the planet.
Namely, this is the case for the 0.25~bar (in the polar regions), 2~bar (in the PR92 scheme) and 4~bar simulations. 
In other words, both microphysics- and cloud-depth based parameterisations diagnose some amount of cloud ice and deep convection on the night side, generally eastward of the evening terminator.
In the high-pressure cases, this is likely due to a combination of the cloud advection by stronger zonal wind in the double-jet circulation regime and a warmer surface triggering convection on the night-side.
The latter demonstrates the importance of including ocean heat transport in future studies \citep[e.g.,][]{DelGenio19_habitable, Yang19_ocean}

\section{Discussion}
\label{sec:discussion}
\subsection{Implications for atmospheric chemistry}
\label{sec:implications_chem}
In this study, we use high-resolution climate simulations with explicit convection to explore the dependence of lightning climatology on the surface air pressure.
Our simulations show a clear trend, suggesting that lightning might be more frequent in thinner atmospheres (\qtyrange{0.25}{1}{\bar}) and less frequent in thicker atmospheres (\qtyrange{2}{10}{\bar}).
This could be further exacerbated by the Paschen law \citep[e.g.,][]{Helling13_ionization, Riousset24_generalized}, which states that for a given distance between charged particles the breakdown voltage that needs to be exceeded to start an electric discharge is proportional to pressure.
In other words, tropospheric pressure in atmospheres more massive than Earth's may prevent electric discharges from happening even when meteorological conditions are met.
At the same time, the amount of energy released per flash would be higher, meaning that lightning in a high-pressure atmosphere would be rarer but stronger.
Therefore, a balance between the total flash rate and total energy needs to be estimated in future studies.

The key implication of our study is that lightning-induced chemistry could be more prevalent in thinner atmospheres, with quantitative estimates depending on the lightning scheme used in the model.
The chemical impact fundamentally depends on the ambient atmospheric composition \citep{Chameides81_rates, NnaMvondo01_production, Ardaseva17_lightning, Harman18_abiotic, Barth24_effect}.
For oxygenated atmospheres on tidally locked exoplanets, \citet{Braam22_lightning-induced} find lower flash rates in simulations of Proxima Centauri b using the PR92 scheme (see also Section~\ref{sec:res_pr92}) and report no significant impact of lightning-induced nitrogen oxides (NO$_\mathrm{x}$) on the ozone column density.
On the other hand, \cite{Luo23_coupled} use prescribed surface NO$_\mathrm{x}$ emissions and find oscillations in the ozone column density, which would constitute a biosignature.
However, other works have shown that oxygen and ozone can accumulate abiotically on exoplanets \citep[e.g.,][]{Hu12_photochemistry, Domagal-Goldman14_abiotic, Tian14_high, Harman18_abiotic}.
The higher flash rates that we find can enhance NO$_\mathrm{x}$ abundances and present an alternative to the biosignature scenario in \cite{Luo23_coupled}.
Whilst a coupled chemistry-climate simulation with the enhanced flash rates is beyond the scope of this study, we can provide a first-order quantitative estimate.

\cite{Luo23_coupled} note that NO$_\mathrm{x}$ emissions due to biological nitrogen fixation are 80 times those due to lightning (\qty{8}{\nflux} versus \qty{0.01}{\nflux}).
In \citet{Braam22_lightning-induced}, the lightning-induced NO$_\mathrm{x}$ delivers \qty{0.035}{\nflux}, with global mean LFR of \qty{2.3e-3}{\lfr}.
While for a slightly different planetary setup, this rate is ${\sim}$6 times lower than our experiments with the 1~bar surface pressure predict using the same PR92 scheme (\qty{1.4e-2}{\lfr}, see Fig.~\ref{fig:lfr_pr92_maps}c).
The M09 scheme predicts an even higher global mean LFR of \qty{3.8e-1}{\lfr} (Fig.~\ref{fig:lfr_m09_maps}c), which is ${\sim}27$ times that of the PR92 scheme.
Combined, NO$_\mathrm{x}$ emissions can be up to 162 times higher than those reported by \citet{Braam22_lightning-induced}, bringing the nitrogen fixation by lightning up to \qty{5.67}{\nflux}, much closer to the biological emission fluxes of \qty{8}{\nflux} used by \cite{Luo23_coupled}.
Therefore, the enhanced lightning-induced chemistry may be able to mimic the oscillations in ozone column density abiotically.
Future work should be directed to testing these predictions in high-resolution coupled chemistry-climate simulations.

Dayside photochemistry is likely to play a major role in the reported dayside-nightside asymmetries on tidally locked exoplanets \citep[see Fig.~8 in][]{Braam22_lightning-induced}.
Both Fig.~\ref{fig:lfr_m09_maps} and Fig.~\ref{fig:lfr_pr92_maps} show that a significant asymmetry remain in the hemispheric mean LFR.
This suggests that the asymmetric distribution --- simple species predominantly on the dayside, more complex species on the nightside --- should broadly hold.
If hydrocarbons are present in the atmosphere, enhanced NO$_\mathrm{x}$ emissions may promote ozone formation in the troposphere through the photochemical smog mechanism \citep{Haagen-Smit53_ozone}.
With regards to the reducing atmospheres of early Earth and exoplanets, enhanced lightning flash rates may promote the production of prebiotically relevant species \citep{Ardaseva17_lightning, Barth24_effect}.
This presents exciting connections to any potential surface reservoirs \citep{Pearce22_rna}.

If prebiotic compounds need shielding from ultra-violet irradiation \citep[e.g.,][]{Ranjan16_influence}, high levels of ozone \citep{Cooke24_lethal} or stellar flares \citep[e.g.,][]{Chen21_persistence, Ridgway22_3d}, they require either dayside-nightside transport \citep{Braam23_stratospheric} or a formation process in the absence of UV radiation.
The latter can be provided if lightning can occur on the night side of the planet, which is in fact predicted for certain surface pressures by both parameterizations used in our study.
Furthermore, a dynamic ocean \citep[e.g.,][]{DelGenio19_habitable} or a higher-order spin-orbit resonance \citep{Turbet16_habitability, Boutle17_exploring, Braam25_earthlike} can result in lightning on the night-side of the planet, potentially affecting the emergence of life.

\subsection{Uncertainties in exoplanet lightning modelling and directions for future work}
\label{sec:caveats}
The main uncertainty in our study stems from the coefficients used in the lightning parameterizations (see Sec.~\ref{sec:lightning_param}).
However, they have been successfully evaluated for Earth weather and climate \citep[e.g.,][]{Wilkinson17_technique, Field18_simulated, Luhar21_assessing} so we expect them to perform reasonably well for a temperate climate with water as a condensible species.
Moreover, both parameterizations are rooted in the fundamental physics of atmospheric electricity and are linked to the key meteorological variables associated with thunderstorms, i.e. ice content \citep{Han21_cloud} and cloud height \citep{Vonnegut63_facts}.
Therefore, our results provide a qualitative assessment of lightning activity for different surface pressures.
A possible route towards a more quantitative result is to use an explicit electric charging scheme coupled with a cloud-resolving model \citep[e.g.,][]{Fierro13_implementation}.
In the absence of in-situ observations, this may be the best way of tuning convection- or microphysics-based parameterisations in climate models for different planetary atmospheres.

A key unknown in lightning modelling for terrestrial exoplanets is the bulk composition of the atmosphere.
It determines what molecules are created by lightning \citep{Barth24_effect} and, via changes in the thermodynamics of the atmosphere, determines where and how frequent convective lightning storms are across the planet.
A feedback mechanism can also reinforce this.
Earth studies have suggested a positive feedback between lightning and \ce{O3} \citep{Finney18_projected}.
An increase in lightning activity in a warmer climate leads to an increase in tropospheric \ce{O3}, which by absorbing infrared radiation causes further warming and higher frequency of thunderstorms; however this effect can be counteracted by a decrease in another greenhouse gas, \ce{CH4}, which is depleted by an increase in \ce{O3}.

In this study, we adopted the same bulk composition across our six simulations.
A systematic mapping of the parameter space in terms of bulk atmospheric composition will be conducted in a follow-up study.
Preliminary simulations with higher levels of \ce{CO2} (not shown), indicate a non-linear dependence of the lightning precursors on the amount of these greenhouse gases.
For example, we find that in the \ce{CO2}-dominated THAI Hab~2 simulation \citep{Fauchez20_thai_protocol, Sergeev22_thai}, the convective cloud depths are substantially higher than those in our reference case (\ce{N2}-dominated, 1~bar), making the PR92 scheme to predict an order of magnitude higher LFR.
On the other hand, the amount of total cloud ice is similar to that in our reference case, resulting in the PR92 estimate similar to that made by the M09 scheme.
This shows that the PR92 scheme is less applicable for exoplanet simulations: this scheme was derived for convective clouds on Earth, with a freezing level at a particular height derived from climatology.
Microphysics-based lightning schemes, such as M09, are more flexible and could be easier adapted to an extraterrestrial climate or even a warmer climate on Earth.
For example, \cite{Finney18_projected} note that more robust lightning projections for a changing climate can be done with parameterizations based on cloud ice and microphysical processes. 

Another key factor affecting lightning activity is the land-ocean distribution.
On Earth, continental areas and islands experience substantially more lightning than ocean areas \citep[e.g.,][]{Christian03_global, Field18_simulated}.
Compared to the ocean, land has lower heat capacity, making the lower atmosphere more unstable to vertical motions which are crucial for deep convection, hydrometeor collisions, charge separation, and lightning \citep{Williams02_physical}.
While beyond the scope of the present study, our preliminary experiments indicate that placing a continent on the day side of the planet indeed changes the total LFR.
The sign of this change, however, depends on the lightning parameterisation.
The M09 scheme produces fewer lightning flashes compared the aquaplanet reference case, while the PR92 scheme produces more.
Increasing the size the continent causes the total LFR to fall or rise depending on the parameterisation, which indicates that a substellar continent leads to stronger vertical motions and taller clouds but at the same time dries the atmosphere and reduces the cloud ice amount in the troposphere \citep{Lewis18_influence}.

In the context of no-continent (aquaplanet) modelling, chemical properties of sea surface water add another layer of complexity.
Laboratory studies suggested that lightning frequency is correlated to the ocean alkalinity \citep{Silverman21_possible}, potentially explaining the prevalence of superbolts over the oceans \citep{Holzworth19_global}.
Furthermore, high ocean salinity increases the strength of individual lightning strikes because of the higher conductivity of saltier water \citep{Asfur20_why}.
Ocean water properties on exoplanets could be expected to deviate a lot from their typical values on Earth \citep[e.g.,][]{Cullum16_salinity, Olson22_effect}, so while the LFR on an aquaplanet may be small \citep{Braam22_lightning-induced}, its overall impact may be larger than that on Earth.
For example, a more massive atmosphere may have weaker precipitation \citep{Xiong22_smaller} and thus higher salinity of the surface sea water, which may result in less frequent but more intense lightning strikes.
More laboratory studies are required to constrain this effect.

For tidally locked planets, the prevalence of thunderstorms around on the day side, even over the wide range of pressures from 0.25 to 10~bar, also presents an efficient way to model lightning-induced chemistry.
Instead of a \textit{global} storm-resolving setup \citep[used here and in e.g.,][]{Yang23_cloud}, it is more computationally cheap to simulate lightning in a limited-area domain using either a nesting suite \citep{Zhang17_surface, Sergeev20_atmospheric} or a locally refined mesh \citep{Bindle21_grid-stretching, Gu24_simulated, Sergeev24_impact}.
We will apply the latter technique to study lightning on tidally locked planets in our future work.

\section{Conclusions}
\label{sec:conclusions}
We used a state-of-the-art storm-resolving 3D model to simulate lightning activity on a terrestrial tidally locked planet with varying amount of atmospheric mass, equivalent to the surface pressure ranging from 0.25 to 10~bar.
Our key findings are as follows.
\begin{enumerate}
    \item Lightning flash rate monotonically decreases with surface pressure, as estimated both by a microphysics-based and cloud-depth based parameterisations.
    \item With increasing surface pressure, lightning activity reduces due to a weaker convection and warmer climate.
    Convection is weaker because the moist adiabatic lapse rate increases with pressure.
    Weaker convection, together with a stronger greenhouse effect and a change in the global circulation, result in shallower convective clouds with less cloud ice in our high-pressure simulations.
    \item The microphysics-based scheme produces roughly an order of magnitude more lightning than the cloud-depth based scheme.
    Compared to the cloud depth approach, microphysics-based schemes offer a more physically-motivated and flexible approach to modelling lightning on exoplanets.
    \item The key climate variables such as the global mean surface temperature, cloud ice content, and precipitation show monotonic dependence on the surface pressure.
    \item The general circulation of the atmosphere changes from the single-jet regime for low-pressure climates to the double-jet regime for high-pressure climates, which affects the spatial distribution of deep convection and the resulting lightning occurrence.
    In some high-pressure simulations, the model produces lightning on the night side eastward of the evening terminator.
    This may affect our theoretical expectations of atmospheric chemistry on tidally locked exoplanets, although the dayside-nightside asymmetry in subsequent photochemistry remains.
\end{enumerate}

\section*{Acknowledgements}
We are grateful to Dorian Abbot for reviewing this paper.
We thank Yoav Yair for a helpful discussion about the implications of the pressure dependence of lightning.
We thank Jonathan Wilkinson and Paul Field for their help with setting up the UM lightning and microphysics schemes.
We thank Nathan Mayne for supporting LW's undergraduate summer research internship on which this study is based.
JWM was supported by an EPSRC Vacation Internship. 
DES was supported by UKRI Future Leaders Fellowship (\texttt{MR/T040866/1}).
JKE-N would like to thank the Hill Family Scholarship.
The Hill Family Scholarship has been generously supported by University of Exeter alumnus, and president of the University's US Foundation Graham Hill (Economic \& Political Development, 1992) and other donors to the US Foundation.
MB appreciates support from a CSH Fellowship.
We acknowledge use of the Monsoon2 system, a collaborative facility supplied under the Joint Weather and Climate Research Programme, a strategic partnership between the Met Office and the Natural Environment Research Council.

\section*{Software}
Material produced using Met Office Software.
The Met Office Unified Model is available for use under licence; see \url{http://www.metoffice.gov.uk/research/modelling-systems/unified-model}.
Scripts to post-process and visualise the model data are available on GitHub: \url{https://github.com/dennissergeev/thunderstruck} and are dependent on the following open-source Python libraries: \texttt{aeolus} \citep{Sergeev24_aeolus}, \texttt{iris} \citep{iris}, \texttt{jupyter} \citep{Kluyver16_jupyter}, \texttt{matplotlib} \citep{Hunter07_matplotlib}, \texttt{numpy} \citep{Harris20_array}.

\section*{Data Availability}
The UM dataset used in this study is available on Zenodo: \\ \url{https://doi.org/10.5281/zenodo.15190801}. 
\bibliographystyle{mnras}
\bibliography{references}


\bsp	
\label{lastpage}
\end{document}